# Dependence of band gaps in *d*-electron perovskite oxides on magnetism

Julien Varignon[1], Oleksandr I. Malyi[2], and Alex Zunger[2]

[1]Laboratoire CRISMAT, CNRS UMR 6508, ENSICAEN, Normandie Université, 6 Boulevard Maréchal Juin, F-14050 Caen Cedex 4, France
[2]Renewable and Sustainable Energy Institute, University of Colorado, Boulder, Colorado 80309, USA

Understanding the controlling principles of band gaps trends in *d* electron perovskites is needed both for gauging metal-insulator transitions, as well as their application in catalysis and doping. The magnitude of this band gap is rather different for different magnetic spin configurations. We find *via* electronic structure theory that the factors that connect gapping magnitudes to magnetism depend on the nature of the band edge orbital character (BEOC) and surprisingly scale with the number of antiferromagnetic contacts $z_i$ between neighboring transition metal ions. The dependence is weak when the BEOC are (*d,d*)-like ('Mott insulators'), whereas this dependence is rather strong in (*p,d*)-like ('charge transfer' insulators). These unexpected rules are traced to the reduced orbital interactions through the increase in the number of antiferromagnetic contacts between transition metal ions. The impact of magnetic order is not limited to the band gap magnitude and includes also the magnitude of lattice distortions connected to the electronic structure. These results highlight the importance of establishing in electronic structure theory of gap-related phenomena (doping, transport, metal-insulator transitions, conductive interfaces) the appropriate magnetic order.

Oxide perovskites ABO$_3$ have long attracted broad interest thanks to their diverse range of properties encompassing ferroelectricity, magnetism and superconductivity [1]. The band gap of perovskites stands out as one of its key quantities used not only to gauge metal-insulator transitions [2] and the physics of correlations [3], but also as a basic metric for designing catalysis, doping, transparent conductors, conductive interfaces and switching of ferroelectrics. Thus, the ability of electronic structure theories to correctly track the band gaps of such solids has been an important challenge. Perhaps the effect that held the community in most fascination in this regard has been the persistent existence of insulating band gaps in long range ordered (LRO) perovskite crystal structures where simplified band theoretic view would suggest a metallic state when the number of electrons per primitive cell is odd [4–6]. This puzzle motivated the strong correlation physics approaches, encoded *e.g.*, in the celebrated Mott-Hubbard model [7,8] as a unifying gap-opening mechanism. Yet, it has been also known [9] that cell-doubling anti-ferromagnetic (AFM) long range spin order – a *spin symmetry breaking* describable by mean-field like band theory – opens band gaps in otherwise metallic, low temperature ground state phases of *d* electron perovskites. More recently, different modes of *spatial symmetry breaking* (illustrated in Supplementary 1) were shown to explain gapping in such long range ordered systems as well as in high temperature paramagnetic cell lacking long range spin-order [10–16].

These developments in the understanding of gapping mechanisms in *d* electron oxides by symmetry breaking [10–16], face, however, the yet unexplained occurrence of large systematic variations of band gaps with magnetic order illustrated via first principles calculations in Fig.1. Trends in electronic properties of the particular series of Lanthanides LaBO$_3$ (B=Sc to Cu) were discussed by He and Franchini [17], yet the role of magnetic order on the band gap magnitudes remains unknown. In ABO$_3$ perovskites based on corner sharing BO$_6$ motifs, these magnetic orders can be classified as a function of the number z of AFM contacts between two 1$^{st}$ neighbors (z$_1$) and up to the 3$^{rd}$ (z$_3$) nearest neighbor TM sites (see insert of Fig.1 for definitions of z$_1$, z$_2$ and z$_3$ and Supplementary SI2 which provides illustrations of the possible magnetic configurations). We see in Fig.1 that the dependence of the insulating gap on magnetic configuration is rather weak in Mott compounds – where both valence band maximum (VBM) and conduction band minimum (CBM) have dominant *d* orbital character [denoted as (*d,d\**) band edges], whereas this dependence is rather strong in charge transfer insulators i.e. compounds with (*p,d\**) band edges. Here we show that such trends in the magnitudes of insulating band gaps of ordered and disordered spin configurations scale in a surprising manner with the number $z_i$ of antiferromagnetic contacts between i-th neighboring transition metal sites. This result is explained by the reduction in orbital interactions between neighboring TM sites through antiferromagnetic couplings. This realization clarifies how magnetism controls band gaps in perovskites having different band edge orbital character. Furthermore, the same physics explains also the effect of magnetism on the orbital localization and structural distortions such as Jahn-Teller and bond disproportionation. The importance of understanding such trends lies in the ability to control optical response, doping, catalysis and band alignment at oxide interfaces by adjusting the band gap amplitude.

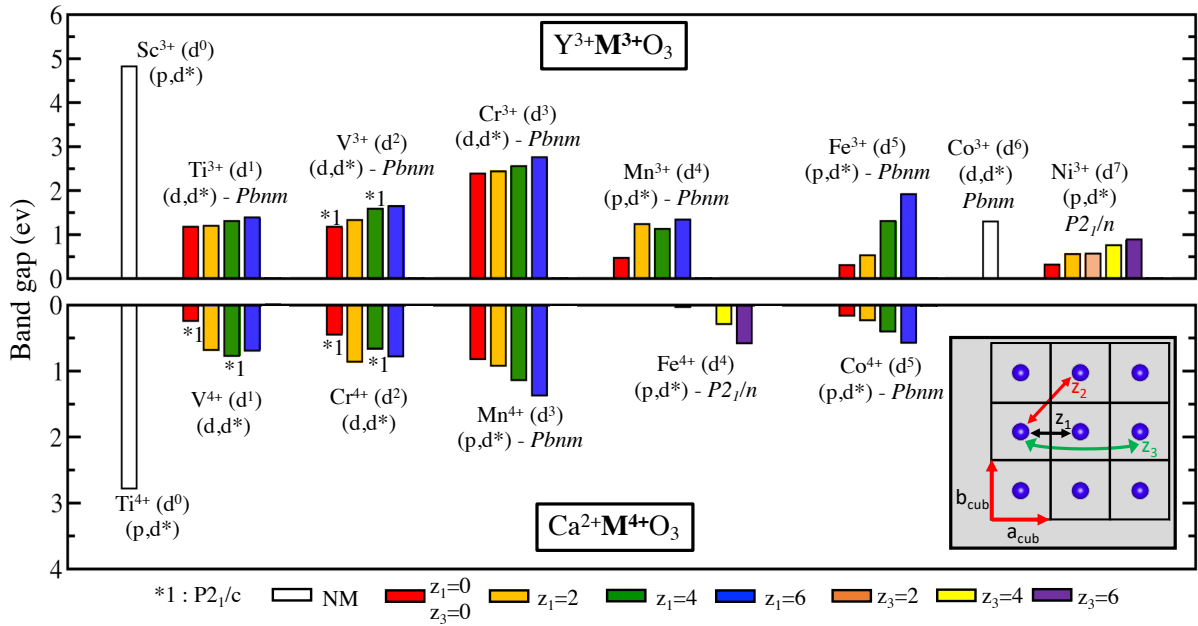

*Figure 1:* *Calculated DFT minimal energy band gaps (in eV) displayed by ABO$_3$ perovskites as a function of the magnetic order described by the number zi of AFM contacts. These include FM ($z_1$=0), AFMA ($z_1$=2), AFMC ($z_1$=4), AFMG ($z_1$=6) orders. NM stands for a non-spin polarized calculations. The compounds shown have B atom electron count d$^n$, formal oxidation state being either in a 3+ (upper panel) or a 4+ (lower panel) and crystal space groups Pbnm except CaFeO$_3$ and YNiO$_3$ that adopt the P2$_1$/n symmetry and YVO$_3$, CaCrO$_3$ and CaVO$_3$ that adopt a P2$_1$/c symmetry. $z_1$, $z_2$ and $z_3$ are defined in insert assuming a hypothetical high symmetry cubic cell for clarity. We use an on-site energy U=3 eV for all cases, avoiding fitting of individual cases. This U value might be overestimated for CaVO$_3$ that we find insulating while it is a metal experimentally. A lower U value was previously fitted to 1.25 eV [10].*

**Some elements of the method:** DFT calculations were performed with the GGA PBEsol exchange-correlation functional [18]. To better cancel the spurious self-interactions errors inherent to approximate DFT [19], we apply an on-site U potential of 3 eV, acting on the *3d* orbitals [20]. No attempt is made here to optimize this constant U potential value for individual compounds, as we are interested here in general metal *vs* insulator trends and overall scaling of insulating gaps with physical descriptors such as the number of AFM contacts rather than fitting experiment [21]. Crystallographic cells used in the calculations correspond to orthorhombic *Pbnm* symmetry (4 formula units/cell) allowing the $a^+a^+c^-$ octahedra rotations in Glazer's notation [22] ($YScO_3$, $YTiO_3$, $YCrO_3$, $YFeO_3$, $YCoO_3$, $CaTiO_3$, $CaMnO_3$, $CaCoO_3$) or the lower *P2₁/c* space groups ($YVO_3$, $CaVO_3$, $CaCrO_3$) and *P2₁/n* ($CaFeO_3$, $YNiO_3$) symmetries allowing the Jahn-Teller $Q_2^-$ or bond disproportionation $B_{oc}$ distortion modes (see Figure SI1 for sketches of these distortions). Several spin-orders are tested and are described by the number of AFM contacts couplings $z_i$ between 1$^{st}$ ($z_1$) and 3$^{rd}$ ($z_3$) nearest neighbor (nn) TM sites (see insert of Fig 1 and supplementary SI 2 for illustrations of magnetic orders). Interactions may be restricted to 1$^{st}$ nn spins in most compounds, yielding the usual FM ($z_1$=0), AFMA ($z_1$=2), AFMC ($z_1$=4) & AFMG ($z_1$=6) orders. Yet some materials can show two magnetic sublattices due to disproportionation effects and thus have 3rd nearest interactions, yielding the AFME ($z_3$=4) and AFMT ($z_3$=6) orders. Compounds with $z_3$=2, 4 or 6 require larger supercells, corresponding to (1,1,2), (2,1,1) and (2,1,2) orthorhombic supercells, respectively. Further details on the calculations can be found in Supplementary information 3 and 4.

***Classification of d electron oxide perovskites according to the orbital character of their band edges:*** *d* electron oxides were traditionally [23,24] classified into 3 groups based on their band edge orbital character. These classes include (i) compounds where both the VBM and CBM are made of the B atom *d*-orbitals (*d,d\**) – historically termed "Mott type compounds"; (ii) compounds with VBM and CBM band edges made of anion *p* and cation *d* orbitals (*p,d\**), respectively – historically termed as "positive charge transfer state". Here the occupied band includes a trapped metal electron such as $Ti^{4+}+e=Ti^{3+}$ also known as "electron polaron"; (iii) (*p,p\**) compounds where the VBM and CBM are made of *p* orbitals – historically termed "negative charge transfer" compounds. Classification (i)-(iii) above complete the standard BEOC common in ordinary insulators of (*p,s\**) band edges such as main group oxides and chalcogenides.

The classification noted above is based on semi empirical approaches (such as tight binding models [24,25]), where the chemical trends are not always faithfully reproduced. To characterize the different band edge orbital types self consistently from DFT we have compute the charge transfer energy $\Delta E_{dp}$ – the energy difference between TM-$d$ states and anion O $p$ states – using Wannier Functions (WFs) built from our DFT Bloch band structure. We use for this purpose the low temperature stable orthorhombic *Pbnm*, *P2$_1$/c* or *P2$_1$/n* symmetries. Within an orthorhombic cell containing 4 formula units for instance, we thus seek to build 4x5 $d$-like and 12x3 $p$-like WFs, i.e. a total of 56 WFs. To that end, we have projected the 56 Kohn-Sham states located around the Fermi level that always have dominant O $p$ and B $d$-like character on initial atomic orbitals of $p$ and $d$ characters centered on O and TM ions, respectively, in order to extract the initial gauge matrix for the localization procedure. After optimization of the WFs through spread minimization (as implemented in the wannier90 package [26–28]), we have extracted the on-site energy $E_d$ and $E_p$ associated with each WFs of $d$ and $p$-like character. The charge transfer energy ΔE$_{dp}$ is then defined as the difference between the average on-site energy of $d$-like WFs $\overline{E_d}$ and $p$-like WFs $\overline{E_p}$, i.e. $\Delta E_{dp} = \overline{E_d} - \overline{E_p}$.

Figure 2 depicts the calculated Wannier charge transfer energy $\Delta E_{dp}$ as a function of the number of unpaired electrons on the transition metal $d$ orbitals. We see that the sign change of $\Delta E_{dp}$ marking the transition from ($d,d^*$) band edges to ($p,d^*$) band edges shifts systematically to lower values of $d$ electron count upon (i) adding more electrons on $d$ orbitals at a fixed formal oxidation state (FOS) and (ii) for a smaller $d$ electron count for larger FOS (*i.e. $d^4$* for B$^{3+}$ *vs. $d^3$* for B$^{4+}$ cations). Interestingly, a low or high spin state is seen to affect the type of band edge orbital type. We see that YNiO$_3$, with 7 $d$ electrons has an average spin moment of 1 μ$_B$ being in a low spin state, and is seen in Fig.2 to be on the verge of a transition between a Mott insulator and a charge transfer insulators. At the same time, YFeO$_3$, with a smaller number of d electrons but with a high spin state, is clearly a charge transfer insulator. Figure 2 defines a DFT predictor for classifying compounds into different BEOC categories without relying on semi empirical parameterization [24,25].

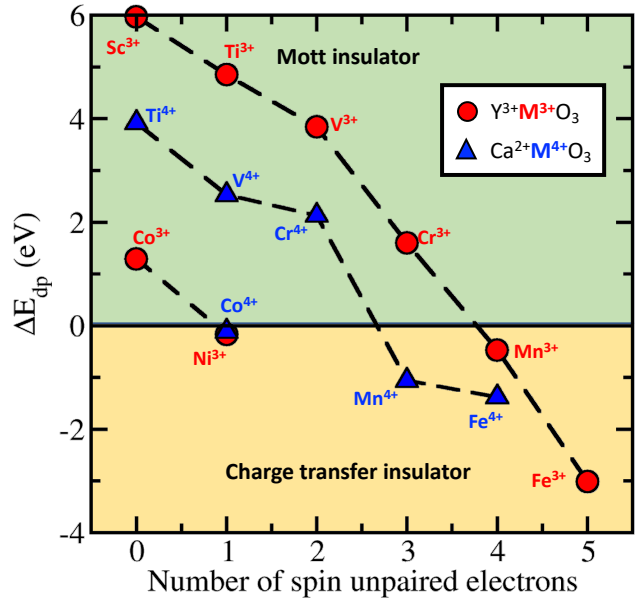

*Figure 2: Wannier charge transfer energy $\Delta E_{dp}$ (in eV) (difference between metal d and anion p-like Wannier functions energies) as a function of the number of unpaired electrons on each B cation in either a 3+ (red filled squares) or a 4+ (blue filled triangles) formal oxidation state. The transition between Mott insulator and charge transfer insulators occurs when $\Delta E_{dp}$ changes sign. We have used the magnetic order with $z_1=6$ for all compounds at the exception of $YTiO_3$ ($z_1=0$), $YNiO_3$ and $CaFeO_3$ ($z_3=6$) and $CaTiO_3$, $YScO_3$ and $YCoO_3$ (Nonmagnetic solution).*

**DFT predicted trends in band gap magnitudes for different band edges and magnetic configurations:** For orthorhombic (*Pbnm*) and monoclinic (*P2$_1$/c* or *P2$_1$/n*) structures we report in Figure 1 the band gap value as a function of the *d* electron count for different number z of AFM contacts between nearest neighbor sites. Several trends, encodable in rules (explained later) emerge:

(i) The FM ($z_1=0$) spin order shows the smallest band gap magnitude for all compounds;

(ii) Compounds where the 1$^{st}$ nn spins are antiferromagnetically coupled (*i.e.* $z_1=$ 6, being AFMG order) have the largest gaps when the structure is orthorhombic *Pbnm,* whatever the magnetic order is ($YTiO_3$, $YCrO_3$, $YFeO_3$, $CaMnO_3$, $YCoO_3$, $CaCoO_3$);

(iii) The gap increases monotonically with increasing number $z_1$ of AFM contacts for compounds adopting LT *Pbnm* symmetry. This is independent of the BEOC;

(iv) Compounds with (*d, d**) band edges (Mott insulators) such as 3+ FOS (YTiO$_3$, YVO$_3$, YCrO$_3$) show a rather weak increase in gap with the number $z_1$ of AFM contacts between first nearest neighbor sites. The slope is large for (*p, d**) charge transfer compounds such as the late transition metal oxide perovskites YFeO$_3$ and for perovskites with a 4+ FOS (CaMnO$_3$, CaCoO$_3$);

(v) Compounds that can exhibit a lower symmetry (*e.g.* monoclinic structure either *P2$_1$/c* or *P2$_1$/n* rather than the orthorhombic *Pbnm* symmetry) have band gaps that is not maximum for $z_1$=6 but for $z_3$=6 (CaFeO$_3$ and YNiO$_3$), $z_1$=2 (CaCrO$_3$) or $z_1$=4 (CaVO$_3$).

***Understanding the origin of band gap increase with increasing number of AFM contacts in orthorhombic Pbnm structures***: In order to understand how the band gap increases with the number $z_1$ of 1$^{st}$ nn AFM contacts, we consider the projected density of states (pdos) on B *d* and O *p* levels as a function of $z_1$. Figure 3 illustrated this pdos for YFeO$_3$; similar conclusions are drawn for the other materials. Considering a 3D network of ferromagnetically coupled Fe spins (FM states, $z_1$=0), the band gap E$_g$ is small but the bandwidth W associated with the $t_{2g}$ states in the unoccupied manifold is rather large (W= 1.8 eV). Upon adding 1, 2 and 3D AFM contacts, one observes that E$_g$ increases and at the same time W decreases. The largest DFT gap (E$_g$= 1.92 eV) and smallest bandwidth (W= 0.65 eV) are reached for a purely 3D network of AFM interactions between nearest neighbor sites (*i.e.* AFMG, $z_1$=6).

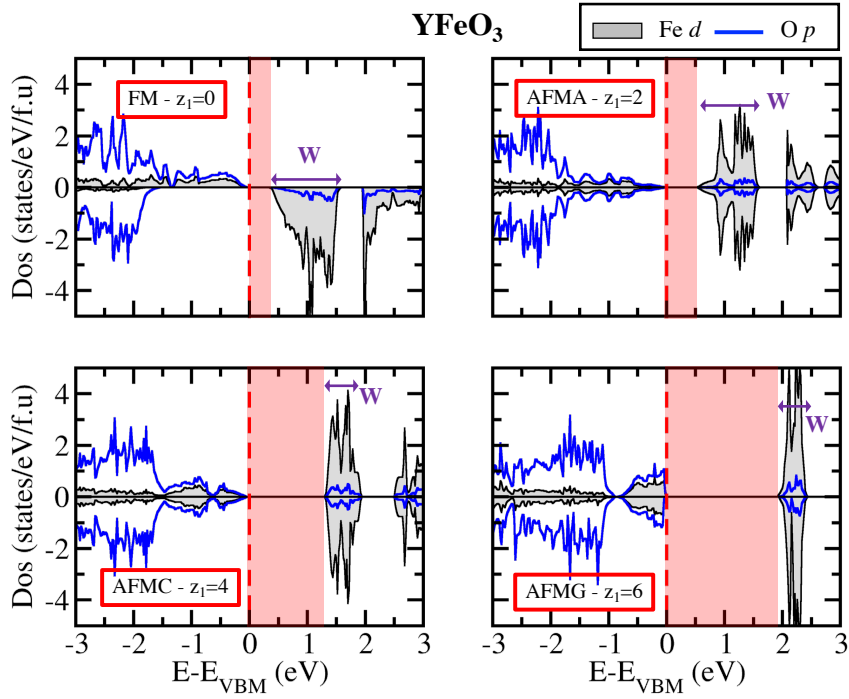

*Figure 3: Projected density of states (in states/eV/f.u) on B d (filled grey) and O p (blue line) states in YFeO₃ for different magnetic orders. The band gap is highlighted by the red area and the bandwidth W of unoccupied $t_{2g}$ states is highlighted by arrows.*

The scaling of gaps with the number of AFM contacts $z_1$ originates from $d$ orbital interactions between nearest neighbor transition metal sites that is related to the transfer integral $t_{dd}$. In the solid, the strength of the orbital interaction between two neighboring sites is directly given by $t_{dd}$ for FM interactions but it is related to $t_{dd}^2/U$ for an AFM interaction, where U is the energy cost of locating two electrons in the same orbital. In practice, $t_{dd}$ is much larger than $t_{dd}^2/U$ and $t_{dd}$ is related to the transfer integral $t_{pd}$ between O $p$ and TM $d$ states through the relation $t_{dd} = t_{pd}^2 / \Delta E_{pd}$ (*i.e.* superexchange mechanism [29]). Orbital interactions thus induce a broadening of the bands either diminishing the band gap (initially given by the exhange splitting strength $\Delta_{ex}$ separating spin up and spin down orbitals locally on each TM sites) by $2t_{dd}$ for a FM configuration, or increasing the band gap by $2t_{dd}^2/U$ for an AFM interaction (See SI 5 for sketches of orbital interactions). Thus, the gap for a given magnetic configuration with $z_1$ AFM contacts between 1st nearest neighbor sites can be written as:

$$E_g(z_1) = \Delta_{ex} - 2(6-z_1)t_{dd} + 2z_1 t_{dd}^2/U \quad (1).$$

It follows from Eq.(1) that

$$E_g(0) = \Delta_{ex} - 12 t_{dd} \quad (2)$$

for a FM solution (*i.e.* $z_1=0$) whereas

$$E_g(6) = \Delta_{ex} + 12 t_{dd}^2/U \quad (3)$$

for an AFMG solution (*i.e.* $z_1=6$).

Figure 4.a shows the gap variation $\Delta E_g$ as a function of the number $z_1$ of AFM interactions introduced in $YTiO_3$, $CaCoO_3$, $YCrO_3$, $CaMnO_3$ and $YFeO_3$. As one can see, the band gap rise upon adding more AFM interactions is similar for all compounds, *modulo* an amplitude that becomes larger when going from Mott insulators to charge transfer insulators ($YTiO_3$ *vs* $CaCoO_3$ or $YFeO_3$, $YCrO_3$ *vs* $CaMnO_3$). This is understandable for $YTiO_3$ ($YFeO_3$) in which the occupied $t_{2g}$ ($e_g$) orbital(s) have a small (large) overlap with O $p$ states thereby producing a weak (large) $t_{pd}$ and thus $t_{dd}$. At fixed magnetic moment but at different FOS such as in $YCrO_3$ and $CaMnO_3$, the effective integral $t_{dd}$ also depends on the relative TM $d$ and O $p$ energy positions $\Delta E_{dp}$ ($t_{dd} \propto 1/\Delta E_{dp}$) [24]. Since $\Delta E_{dp}$ is smaller for $CaMnO_3$ than for $YCrO_3$ (Fig. 2), $t_{dd}$ is thereby larger for $CaMnO_3$.

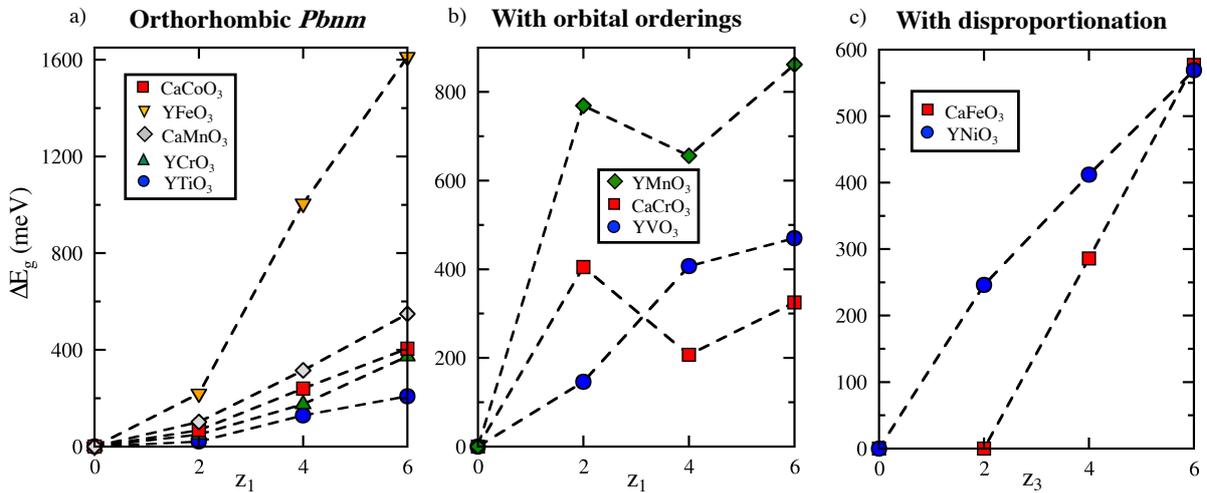

***Figure 4:*** *Band gap variation as a function of the number of AFM contacts $z_1$ in compounds adopting the Pbnm structure (a) and showing the additional $Q_2$ modes (b) and as a function of $z_3$ in materials showing disproportionation effects (c).*

***Additional lattice distortions lowering the orthorhombic Pbnm symmetry induce different orbital occupancies on nearest neighbor.*** As it can be seen in Figure 1, some

compounds deviate from the trend of a monotonic gap evolution with the number of AFM contacts $z_1$ and with a largest value for $z_1=6$. This is the case for $YVO_3$, $CaCrO_3$, $YMnO_3$, $CaFeO_3$ and $YNiO_3$ that show additional and sizable deformation modes with respect to the conventionnal *Pbnm* symmetry (see SI 6 for amplitudes of these modes). These modes are: (i) the bond disproportionation mode $B_{oc}$ in $CaFeO_3$ and $YNiO_3$; (ii) the $Q_2^+$ octahedra deformation mode in $YVO_3$ (AFMA and AFMG orders, *i.e.* $z_1=2$ and 6, respectively), $CaCrO_3$ (AFMA and AFMG orders) and $YMnO_3$ (all magnetic states) and (iii) the Jahn-Teller mode $Q_2^-$ mode for $YVO_3$ and $CaCrO_3$ (FM, AFMC – $z_1=0$ and 4, respectively). Whatever the origin of these modes, electronically induced or structurally triggered [10,13,30–34], different orbital occupancies between neighboring sites emerge, ultimately modifying magnetic interations between TM elements (*e.g.* Goodenough-Kanamori rules [29,35]).

$Q_2^+$ and $Q_2^-$ mode produce 2D and 3D orthogonal orbital occupancies between 1$^{st}$ nn TM sites, respectively – such as alternating $d_x^2$ and $d_y^2$ in $LaMnO_3$ for instance [13]. Thus, orbital interactions between nn TM sites are already minimized with the $Q_2^-$ mode whatever the magnetic order – the gap would be here initially driven by the crystal field splitting producing the orbital-ordering instead of the exchange splitting and also by the number of contacts between TM showing orthogonally occupied orbitals, see SI 5 – whereas an AFM coupling between 1$^{st}$ nn sites showing similar orbital occupancies for the $Q_2^+$ mode can increase the band gap. The mechanism is verified in our calculations for $YMnO_3$ or $CaCrO_3$ that both show a large $Q_2^+$ mode and where AFMA ($z_1=2$) and AFMG ($z_1=6$) orders show the largest band gaps (*i.e.* AFM coupling between 1$^{st}$ nn TM exhbiting similar orbital occupancies) as shown in Figure 4.b. In principle, materials exhibiting the $Q_2^-$ mode should exhibit a gap independent of magnetic orders. However, due to strongly coupled structral-spin-orbital orders, the $Q_2^+$ mode is favored by octahedra rotation and is connected to AFMA and AFMG orders in $YVO_3$ (see Ref. [13]). Thus this material exhibit an AFMG order in the LT phase experimentally [37,38] for which the band gap is maximized – the (*ab*)-plane AFM contacts come from the fact the $d_{xy}$ orbital is occupied on all V sites and thus can also provide a gap increase.

In disporportionating materials such as $YNiO_3$ and $CaFeO_3$, two different magnetic sublattices forming a rocksalt pattern are enabled by the breathing mode $B_{oc}$. Thus, interactions up to 3$^{rd}$ nn spins have to be taken into account. With an AFMG order, 3$^{rd}$ nn interactions are all ferromagnetic and thus it doesn't minimize orbital interactions between TM sites belonging to the same magnetic sublattice. Substantial minimization of band broadening is realized when TM ions of the same magnetic sublattice are AFM coupled ($z_3=6$) in a similar spirit of band broadening minimization when only 1$^{st}$ nn interactions are important. This is again verified with our simulations: $YNiO_3$ and $CaFeO_3$ indeed show the largest gap value for the when the number $z_3$ of AFM contacts between 3$^{rd}$ nn reaches its maximal value of 6 as shown by figure 4.c.

*Magnetic effects on the additional structural distortion amplitudes:* We do not observe noticeable change in the structural properties for materials exhibiting the orthorhombic *Pbnm* symmetry whatever the magnetic order (see SI6). Nevertheless, we find that materials exhibiting a lower *P2$_1$/c* and *P2$_1$/n* symmetry exhibit a variation of the Jahn-Teller distortion $Q_2^-$ or bond disproportionation $B_{oc}$ amplitude with the number of AFM contacts. Using a symmetry adapted mode analysis in order to extract lattice distortions amplitude with respect to a hypothetical high symmetry cubic cell [39,40] (see SI 6), we indeed observe that (i) the $Q_2^-$ mode is only enabled with FM ($z_1=0$) and AFMC ($z_1=4$) orders for $YVO_3$ and $CaCrO_3$ for instance – *i.e.* magnetic orders with a FM coupling between TM ions along one crystallographic direction – with a constant amplitude of 0.051 Å for $YVO_3$ and a slightly varying amplitude of 0.041 Å (FM) and 0.045 Å (AFMC) for $CaCrO_3$ and (ii) the bond disproportionation amplitude varies with the number of 3$^{rd}$ nn AFM contacts in $YNiO_3$ ($CaFeO_3$), going from 0.082 (0.043) Å, 0.089 (0.046) Å, 0.095 (0.061) Å and 0.094 (0.068) Å for $z_1$ and $z_3=0$, $z_3=2$, $z_3=4$ and $z_3=6$, respectively. While the connection between magnetic orders and the Jahn-Teller mode was explained by Kugel and Khomskii [36,41], we clearly observe that once induced, the amplitude of the $Q_2^-$ Jahn-Teller distortion is rather independent on the AFM or FM nature of couplings between TM elements for compounds showing *(d,d\*)* BEOCs. In compounds showing *(p,d\*)* BEOCs such as $CaFeO_3$, the dependance of the additional lattice distortion becomes strongly connected to the number of relevant AFM contacts. Thus, magnetic effects is not restricted to the gap amplitude, but it is also directly related to the distortion amplitude associated with the gap opening process in perovskite oxides [10–13].

In conclusion, the main consequences of magnetism on band gaps are :

(i) Intermediate states in the gap or polaron formation upon doping a perovksite can be missed if a FM state is used due a strong gap underestimation as well as overdelocalized electronic structures;

(ii) Band alignments at oxide interfaces and associated charge transfer might be incorrectly simulated by restricting DFT simulations with a FM order, notably at oxide interface between ($d,d^*$) and ($p,d^*$) BEOCs compounds where the band gap and related lattice distortions may be dramatically underestimated in the charge transfer insulator;

(iii) False gap closure under epitaxial strain or pressure effects can be observed if the incorrect magnetic order is used, notably with a FM order in which the starting gap and related lattice distortions is already way too underestimated in the bulk phase.

All drawbacks related to gap properties might be circumvented by using an appropriate antiferromagnetic order in *first-principles* simulations, notably in charge transfer insulators.

*Acknowledgments*

JV acknowledges access granted to HPC resources of Criann through the projects 2020005 and 2007013 and of Cines through the DARI project A0080911453. Work at CU Boulder was supported under the USA Department of Energy, Basic Energy Science, Office of Energy Sciences, Materials Science and Engineering Division, under grant DE-SC0010467.